\documentclass[aps,prc,superscriptaddress, preprintnumbers, nofootinbib,10pt, showpacs, floatfix]{revtex4-1}

\usepackage{hyperref}

\usepackage{multirow}

\usepackage{feynmf}

\usepackage{amsmath,amssymb}

\usepackage{graphicx,bm}

\usepackage{slashed}

\usepackage{epstopdf}

\usepackage{ulem} 

\usepackage[usenames]{color}

\usepackage{float}

\renewcommand\sout{\bgroup \color{red} \ULdepth=-.5ex \ULset}

 \makeatletter
\newcommand\makebig[2]{%
  \@xp\newcommand\@xp*\csname#1\endcsname{\bBigg@{#2}}%
  \@xp\newcommand\@xp*\csname#1l\endcsname{\@xp\mathopen\csname#1\endcsname}%
  \@xp\newcommand\@xp*\csname#1r\endcsname{\@xp\mathclose\csname#1\endcsname}%
}
\makeatother

\makebig{biggg} {3.5}

\begin{document}
\title{Excluded-volume model for quarkyonic Matter: Three-flavor baryon-quark Mixture}

\author{Dyana C. Duarte}
\email[]{dyduarte@uw.edu}
\author{Saul Hernandez-Ortiz}
\email[]{saulhdz@uw.edu}
\author{Kie Sang Jeong}
\email[]{ksjeong@uw.edu}
\affiliation{Institute for Nuclear Theory, University of Washington, Seattle, WA 98195, USA}

\date{\today}

\preprint{INT-PUB-20-001}

\begin{abstract}
The single-flavor excluded-volume model based on the effective size of baryons reproduces the hard-soft density evolution of the equation of state (EoS) required by the recent studies of GW170817. This phenomenological model basically realizes the concept of quarkyonic matter which is introduced from large-$N_{c}$ gauge theory for dense matter. Enhanced nucleon interactions and dynamically generated quark degrees of freedom can reproduce the hard-soft evolution of the EoS. In this paper, we extend the excluded-volume model to a three-flavor system by considering electromagnetic charge and possible weak equilibrium in order to obtain a proper description for the hard-soft behavior of the EoS inferred from the gravitational waves observations.
\end{abstract} 
\maketitle

\section{Introduction}

Recent observations of GW170817 and subsequent analyses~\cite{TheLIGOScientific:2017qsa, Abbott:2018exr, Fattoyev:2017jql, Annala:2017llu, Vuorinen:2018qzx, Raithel:2018ncd, Most:2018hfd, Tews:2019cap, Tews:2019ioa, Capano:2019eae} provided important clues for understanding dense nuclear matter. The results strongly imply that the equation of state (EoS) at the highest accessible densities should be hard enough to support a $2 M_{\odot}$ state~\cite{Antoniadis, Demorest} .  At densities somewhat lower than the maximum density, the stiffness should be moderated to satisfy $R_{1.4}\leq 13.5~\textrm{km}$ inferred from the tidal deformability observed at the neutron star inspiral~\cite{Abbott:2018exr, Fattoyev:2017jql, Annala:2017llu, Vuorinen:2018qzx, Raithel:2018ncd, Most:2018hfd, Tews:2019cap, Tews:2019ioa, Capano:2019eae}. It seems hard to satisfy both of the constraints. As the nuclear matter density increases, new degrees of freedom are likely to emerge under enhanced energy density of the system. Emergence of new particles usually makes a softer EoS~\cite{Glendenning:1984jr, Knorren:1995ds}, as the created quasiparticles take low momentum phase space and such massive states, like  hyperons  and $\Delta$ isobars, have various decay channels which usually lead to a lower Fermi level~\cite{Brown:1975di, Cai:2015hya}. Therefore, certain repulsive nuclear interactions are required to support the hard EoS in the high density case~\cite{Hebeler:2013nza, Gandolfi:2013baa}.

However, the neutron star radius problem cannot be solved only by repulsive interaction. Actually, we need other physical properties  to solve the problem. The aforementioned observation confined the possible range of tidal deformability at the 90\% confidence level~\cite{TheLIGOScientific:2017qsa, Abbott:2018exr}. Subsequent analyses suggested even more compact neutron stars~\cite{Fattoyev:2017jql, Annala:2017llu, Raithel:2018ncd, Most:2018hfd, Capano:2019eae}, which require softened EoS beyond some intermediate density regime where the hard EoS is supported. By some unknown physical process, the strong repulsion would be turned off or some kind of phase transition to quark matter can be introduced to explain the observation. However, even if one admits this sudden change, there will still remain debates about the signals of such a hypothetical phase transition. As a candidate for a solution, it is worthwhile to consider a quarkyonic-like model~\cite{McLerran:2007qj, Fukushima:2015bda, McLerran:2018hbz, Jeong:2019lhv} which naturally generates a hard EoS.

The concept of quarkyonic matter is based on large-$N_c$ quantum chromodynamics (QCD)~\cite{tHooft:1973alw, tHooft:1974pnl}. In the large-$N_c$ limit, the Debye screening length diverges, $r_{\textrm{Debye}} \sim O( \sqrt{N_c})  \rightarrow \infty$, as the quark loop is suppressed by $1/N_c$~\cite{McLerran:2007qj}. Thus, if there is a large quark Fermi sphere ($T \rightarrow 0$), the quasi-quark states near the Fermi surface whose momenta are distributed in the range of confinement ($\vert \vec{k}^{Q_i}-\vec{k}^{Q_j} \vert <\Lambda_{\textrm{QCD}}$, $\vert \vec{k}^{Q_i}\vert \simeq k^{Q}_F$ where $i,j=\{1,\cdots N_c\}$) will be confined in the baryon-like state through a mechanism similar to vacuum confinement. In this system, the quark wave functions have two distinguished configurations: almost free quarks filled from the low momentum phase space and the confined waves in the baryon-like state. The confined quarks on the surface are aligned to the momentum direction of the baryon-like state ($k_{F}^{B}\simeq N_c k_{F}^{Q}$). These confined states are understood as  Fermi shell distributions of baryon-like states. The transition from nuclear matter to quarkyonic matter occurs at a few times the nuclear matter density $\rho_0$. In the low density regime ($k_{F}^B < \Lambda_{\textrm{QCD}}$), all the quarks are confined in the nucleon as the quarks can be randomly distributed in the range of $\vert \vec{k}^{Q_i}-\vec{k}^{Q_j} \vert <\Lambda_{\textrm{QCD}}$. When $k_{F}^B \sim O(\Lambda_{\textrm{QCD}})$, quarks begin to fully occupy the low momentum phase space. The saturated states form the Fermi sphere of quarks and, by Pauli's principle, the momenta of confined quarks should be larger than the saturated momenta, which leads to the shell-like distribution~\cite{McLerran:2018hbz, Jeong:2019lhv}. In this transition, the total baryon number and energy density are smoothly varying, but the chemical potential of the confined states is suddenly enhanced by large $N_c$ since [$k_{F}^{B}\sim O( \Lambda_{\textrm{QCD}}] \rightarrow N_c k_{F}^{Q}$). This is not a first-order phase transition as the chemical potential (intensive variable) is suddenly enhanced during the transition and smooth energy density and density (extensive variable) are expected~\cite{McLerran:2007qj}. From this point, most of the baryon number increase is taken by the quark degrees of freedom and the shell-like configurations eventually disappear at the extreme density limit [$k^Q_F\sim O(\sqrt{N_c}\Lambda_{\textrm{QCD}})$], as the Debye screening begins to block the confinement process [$r_{\textrm{Debye}} \sim O({N_c}^{0})$].

This concept was applied to describe the hard-soft density evolution of the EoS in previous literature~\cite{McLerran:2018hbz, Jeong:2019lhv}. The shell-like distribution satisfies the aforementioned requirements and leads to a plausible EoS~\cite{McLerran:2018hbz}. If only the hard-soft density behavior of the quarkyonic EoS is considered, some strong mean-field potential expressed in polynomials of $\rho_{N}$ could be introduced. However, such systems have an intrinsic singularity at infinite density and, even if the quark degrees of freedom are allowed, the baryon and quark density simultaneously increase in the physically relevant density regime. This kind of potential cannot lead to a soft enough EoS and the thin shell-like distribution. If one considers the hard-core repulsive interaction as suggested and calculated in Refs.~\cite{Hamada:1962nq, Herndon:1967zza, Kurihara:1984mh, Rischke:1991ke, Kievsky:1992um, Stoks:1994wp,  Wiringa:1994wb, Yen:1997rv, Machleidt:2000ge, Vovchenko:2015vxa, Zalewski:2015yea,  Redlich:2016dpb, Alba:2016hwx, Vovchenko:2017cbu, Vovchenko:2017zpj, Motornenko:2019arp, Ishii:2006ec, Inoue:2016qxt, Nemura:2017bbw, Hatsuda:2018nes, Inoue:2018axd, Park:2018ukx, Park:2019bsz, Sasaki:2019qnh}, the effective scale of the hard-core repulsion can be measured by some effective size of the baryon, where spatial overlap is not allowed by intrinsic singularity. The single flavor excluded volume model for the dense baryon state~\cite{Jeong:2019lhv} dynamically generates the quark degrees of freedom, and the subsequently appearing shell structure provides the hard-soft density evolution of the EoS as required in previous studies~\cite{Fattoyev:2017jql, Annala:2017llu, Vuorinen:2018qzx, Raithel:2018ncd, Most:2018hfd,  Tews:2019cap, Tews:2019ioa, Capano:2019eae, Masuda:2012ed, Kojo:2014rca, Bedaque:2014sqa, Ma:2018qkg, Tews:2018kmu, Kojo:2019raj, Fujimoto:2019hxv}.

In this paper, we extend the excluded-volume model to a three flavor case and include the effect of electrons and overall charge neutrality within the quarkyonic framework. The work is organized as follows. In Sec.~\ref{sec2}, we first model the baryon-quark mixture under the hard-core repulsive mean-field. By requiring charge neutrality and possible $\beta$ equilibrium, we discuss the conditions which determine the existence of strangeness. Also, in this section, we show that hard-core nature and dynamically generated quark degrees of freedom can reproduce the stiff-moderate evolution of the EoS. Finally, in Sec.~\ref{sec3}, we discuss the possible issues based on the multi flavor excluded-volume model and the expected results from the shell structure, which completes the phenomenological modeling of the quarkyonic concept.

\section{Three flavor excluded-volume model and quark degrees of freedom}\label{sec2}

\subsection{Hard-core repulsive interaction and excluded-volume model for three flavors}

\begin{figure}
\includegraphics[width=0.45\textwidth]{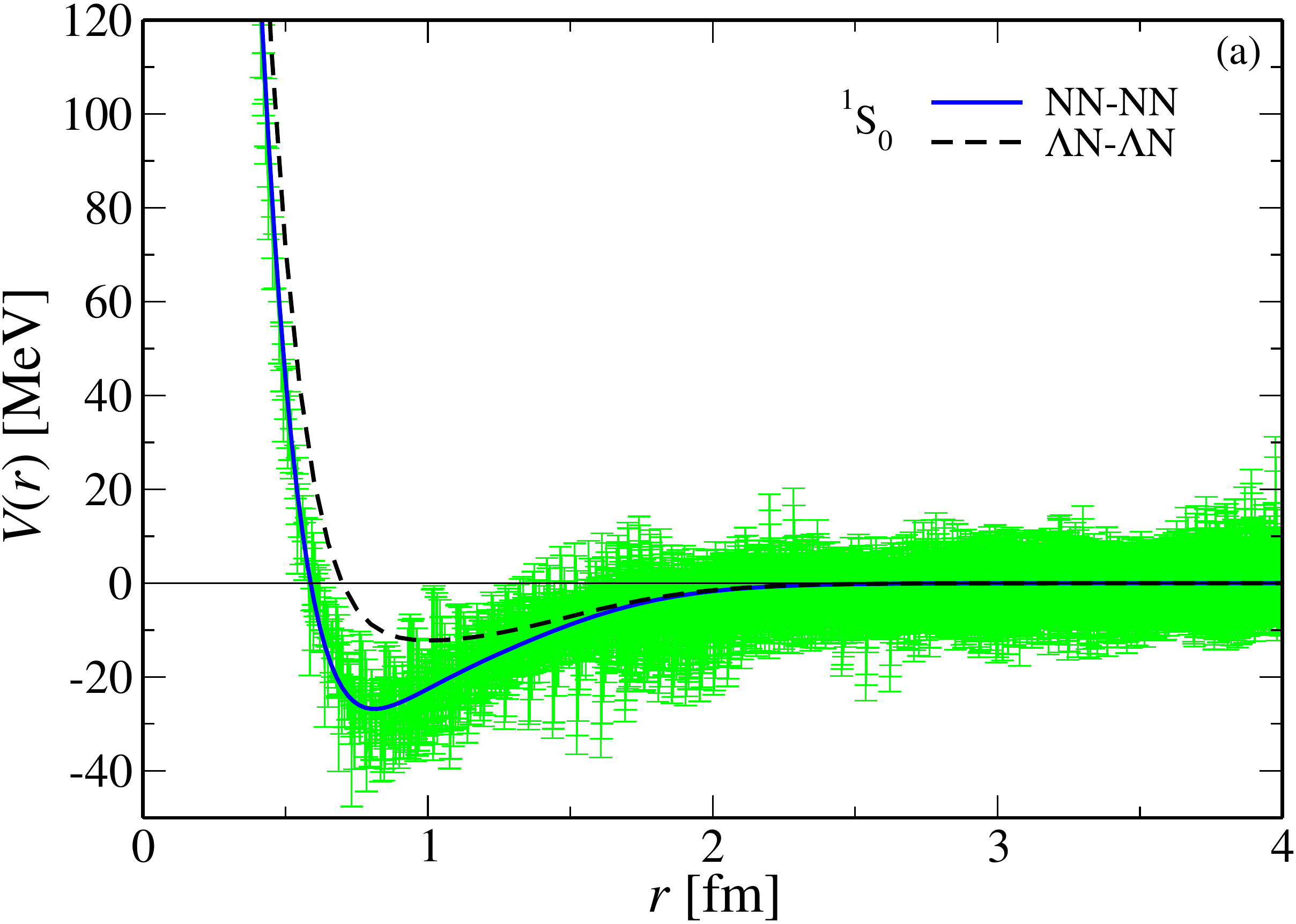}
\includegraphics[width=0.45\textwidth]{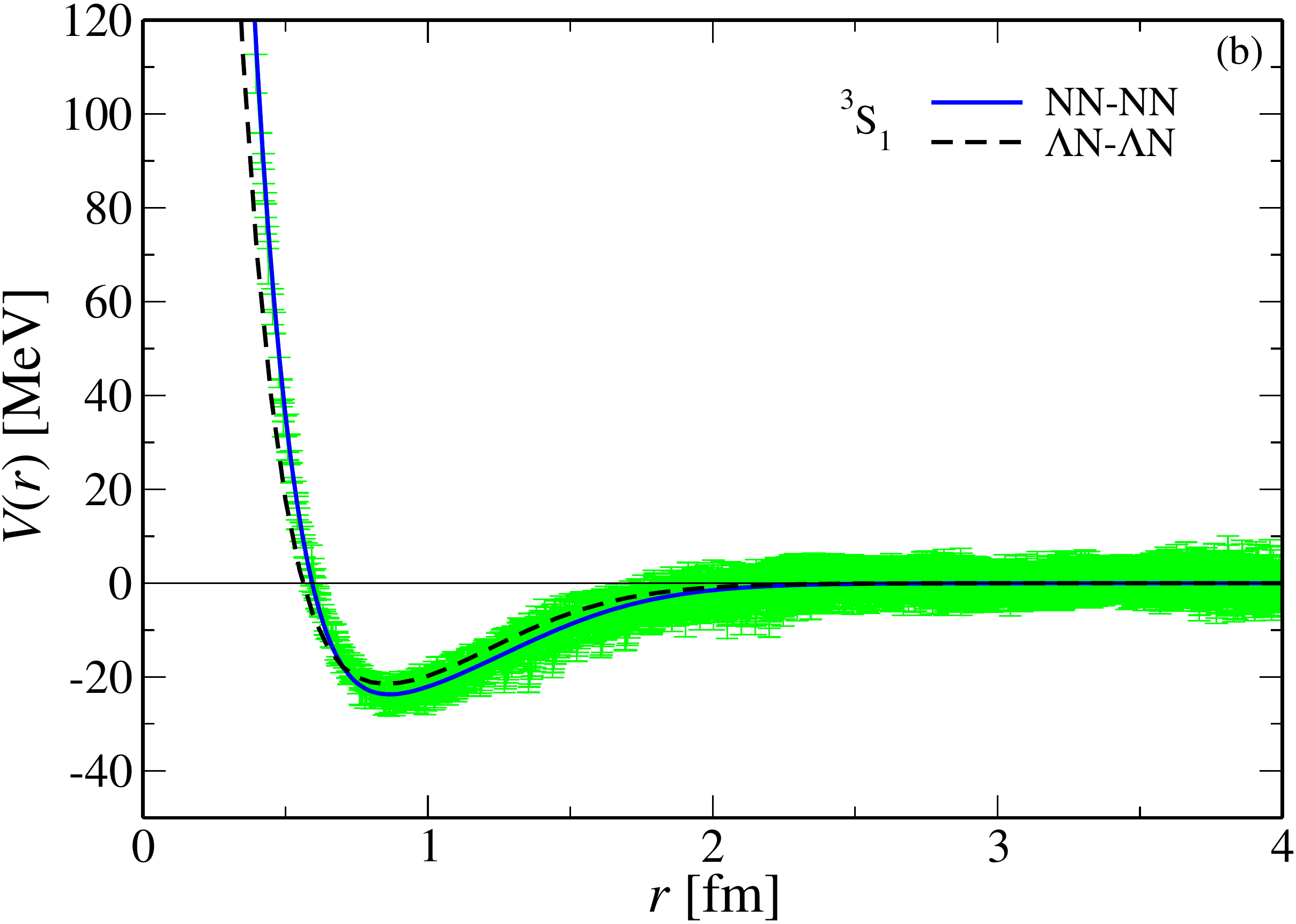}
\caption{$NN$ and $\Lambda N$  potentials quoted from Refs.~\cite{Inoue:2016qxt, Nemura:2017bbw, Hatsuda:2018nes, Inoue:2018axd} under permission. Left (a) and right (b) plots correspond to ${}^{1}S_{0}$ and ${}^{3}S_{1}$ channels, respectively. The black dashed line represents the $\Lambda N$ potential deduced from flavor rotation. The green band represents the uncertainty of the $NN$ potential.} 
\label{fig1}
\end{figure}

The baryon-baryon central potential whose repulsive core appears at the  $0.2-0.4~\textrm{fm}$ scale had been suggested for decades~\cite{Hamada:1962nq, Herndon:1967zza, Kurihara:1984mh, Kievsky:1992um}. As the distance scale of repulsion was compatible with the physical radius of the nucleons, that scale was just considered to be the ideal dense limit where the quark degrees of freedom can appear. However, subsequent studies reported a developed potential whose repulsive core appears around $0.6~\textrm{fm}$ scale~\cite{Stoks:1994wp, Wiringa:1994wb, Machleidt:2000ge}.  This larger scale has been reproduced by the lattice QCD calculation~\cite{Ishii:2006ec, Inoue:2016qxt, Nemura:2017bbw, Hatsuda:2018nes, Inoue:2018axd}, which implies that the effective size due to the repulsion can have a significant dynamical role in the dense regime. The lattice QCD calculation suggested the $S$-wave nucleon-nucleon ($NN$) with short distance singularity in both the spin singlet (${}^{1}S_{0}$) and triplet channels (${}^{3}S_{1}$), as illustrated in Fig.~\ref{fig1}. If one simplifies the central potentials by the infinite well-shaped potential with shallow bounding depth, and defines the hard-core radius where the spatial overlap is not allowed by the infinite potential, the mean distance of the hard-core repulsion would be around $0.6~\textrm{fm}$ ($n_0 \sim 7\rho_0$, where $\rho_0$ denotes the normal nuclear density).  A similar tendency is found from the constituent quark model calculation~\cite{Park:2019bsz}. The $\Lambda$-nucleon ($\Lambda N$) interaction can be deduced by SU(3) flavor rotation, which seems more repulsive than the $NN$ potential in the ${}^{1}S_{0}$ channel and almost same as that in the ${}^{3}S_{1}$ channel.  So the $\Lambda N$ potential can be assumed to have a little bit larger hard-core size than the $NN$ potential. On the other hand, the repulsive potential in the physical world could be different from this simple guess because the error band of the $\Lambda N$ potential~\cite{Inoue:2016qxt, Nemura:2017bbw, Hatsuda:2018nes, Inoue:2018axd} still appears at in non-negligible scale. Moreover, if kaon condensation plays a significant role~\cite{Kaplan:1986yq, Savage:1995kv, Jeong:2016qlk} and the SU(3) flavor symmetry breaking term becomes important, the $\Lambda N$ potential can be reduced. In this work, we consider the possible configurations where the hard-core size of the $\Lambda N$ potential can be larger or smaller than the size of the $NN$ potential, assuming the existence of a hard-core $\Lambda N$ potential. 
 
 We extend the excluded-volume model for the three-flavor ($n$, $p$, and $\Lambda$) case, and each baryon is assumed to have its own effective hard-core size $n_0$ from the repulsive interaction~\cite{Inoue:2016qxt, Nemura:2017bbw, Hatsuda:2018nes, Inoue:2018axd, Park:2018ukx, Park:2019bsz, Sasaki:2019qnh}. This hard-core size can be understood as arising from a repulsive mean-field singular at short distance~\cite{Zalewski:2015yea, Vovchenko:2015vxa, Redlich:2016dpb}. Then, one can redefine the Fermi momentum and number density to include this repulsive nature as follows:
\begin{align}
  n_{B_i}^{ex} &= \frac{n_{B_i}}{1 - n_{\tilde{B}}/n_0}=\frac{2}{(2\pi)^3} \int^{K_{F}^{B_i}}_{0}d^3 k,\label{excdsty}\\
  n_{\tilde{B}} &= n_{n}+n_{p}+(1+\alpha) n_{\Lambda},
\end{align}
where $\alpha$ determines the strength of the hard core repulsive interaction between a surrounding baryon and $\Lambda$ hyperon. The strength is determined in range of $-0.2 < \alpha <0.2$, which is a small variation in comparison with the size used in Refs.~\cite{Alba:2016hwx, Vovchenko:2017zpj}. The $K_{F}^{B_i}$ denoted with a capital letter represents the enhanced baryon Fermi momentum in the reduced available space. In this paper, we only assume the hard-core repulsive interaction to make a simple argument at high density. Following Ref.~\cite{Jeong:2019lhv}, the energy density of the three-flavor baryonic system can be written as
\begin{align}
\varepsilon_{B} & = \left( 1-  \frac{n_{\tilde{B}}}{n_0} \right) \frac{1}{\pi^2} \sum_{i}^{\{n,p,\Lambda\} } \int^{K_{F}^{B_i}}_{0} dk k^2  \left(m_{B_i}^2+k^2 \right)^{\frac{1}{2}}+ \frac{(3 \pi^2)^{\frac{4}{3}}}{4\pi^2}n_{e}^{\frac{4}{3}}\nonumber\\
&\simeq \sum_{i}^{\{n,p,\Lambda\} } \left( \frac{ (3 \pi^2)^{\frac{5}{3}}}{10\pi^2 m_{i} } \frac{n_{i}^{\frac{5}{3}}}{ \left(1- n_{\tilde{B}} /n_0 \right)^{\frac{2}{3}} } + m_{i}n_{i} \right) +\cdots+ \frac{(3 \pi^2)^{\frac{4}{3}}}{4\pi^2}n_{e}^{\frac{4}{3}},
\end{align}
where the electron mass is suppressed and the nonrelativistic limit is taken in last line.  The chemical potential for each baryon can be expressed as~\cite{Jeong:2019lhv} 
\begin{align}
\mu_i =\frac{\partial \varepsilon_{B}}{\partial n_{i}}  =&~ \left( 1- \frac{  n_{\tilde{B}}}{n_0} \right) \left\{ \frac{  {K_F^{B_i}}^2}{\pi^2} \left( {K_F^{B_i}}^2 + m_{B_i}^2 \right)^{\frac{1}{2}}\frac{\partial  K_F^{B_i}}{\partial n_{B_i}}  +   \sum^{\{n,p,\Lambda\}}_{j \neq i} \frac{  {K_F^{B_j}}^2}{\pi^2}\left({K_F^{B_j}}^2 + m_{B_j}^2 \right)^{\frac{1}{2}}  \frac{\partial  K_F^{B_j}}{\partial n_{B_i}}   \right\} \nonumber\\
&~-\frac{\omega_{i}}{n_0}\sum_{l}^{ \{n,p,\Lambda\}} \int_{0}^{{K_F^{B_l}}} dk k^2  \left(m_{B_l}^2+k^2 \right)^{\frac{1}{2}}\nonumber\\ 
=&  \left(  \frac{n_0 -(n_{\tilde{B}}-\omega_i n_{B_i})}{n_0 - n_{\tilde{B}}} \right) \left({K_F^{B_i}}^2  + m_{B_i}^2 \right)^{\frac{1}{2}}  \nonumber\\
&~+  \frac{\omega_i}{n_0}\left\{ \sum_{j \neq i}^{\{n,p,\Lambda\}}   \bar{n}_{B_j}^{ex}\left( {K_F^{B_j}}^2  + m_{B_j}^2 \right)^{\frac{1}{2}} -\sum_k^{\{n,p,\Lambda\}} \frac{1}{\pi^2} \int^{k^{B_k}_F}_{0} dk k^2  \left(k^2+m_{B_k}^2 \right)^{\frac{1}{2}} \right\},\nonumber\\ 
 \simeq &~m_i +  \frac{ (3 \pi^2)^{\frac{5}{3}}}{10 \pi^2 m_{i} } \frac{5}{3} {n_{B_i}^{ex}}^{\frac{2}{3}} + \omega_{i}  \sum_{j  }^{\{n,p,\Lambda\}} \frac{ (3 \pi^2)^{\frac{5}{3}}}{10 \pi^2 m_{j}} \frac{2}{3 n_0} {n_{B_{j}}^{ex}}^{\frac{5}{3}}+\cdots,\label{hchemp}
\end{align}
where the nonrelativistic limit is taken in the last line and $\omega_{i}=\partial n_{\tilde{B}}  / \partial n_{i}$ ($\omega_{n,p}=1$, $\omega_{\Lambda}=1+\alpha$). The partial derivatives are calculated as
\begin{align}
\frac{\partial K_F^{B_i} }{\partial n_{B_i} }=&~\frac{\pi^2}{{K_{F}^{B_i}}^2} \left( 1- \frac{n_{\tilde{B}}}{n_0} \right)^{-2}\left( 1-  \frac{ \left(n_{\tilde{B}}-\omega_i n_{B_i}\right)}{ n_0} \right),\\
\frac{\partial K_F^{B_j} }{\partial n_{B_i} }=&~\frac{\pi^2}{{K_{F}^{B_j}}^2} \left( 1- \frac{n_{\tilde{B}}}{n_0} \right)^{-2} \left( \frac{\omega_{i}  n_{B_j}} {n_0} \right).
\end{align} 
Considering the presumed hard-core size of the particle, this approach can be understood as the cold-dense limit of the van der Waals (vdW) model in Fermi-Dirac statistics~\cite{Rischke:1991ke, Kievsky:1992um, Stoks:1994wp,  Wiringa:1994wb, Yen:1997rv, Machleidt:2000ge, Vovchenko:2015vxa, Zalewski:2015yea, Redlich:2016dpb, Alba:2016hwx, Vovchenko:2017cbu, Vovchenko:2017zpj, Motornenko:2019arp}.\footnote{If one follows Ref.~\cite{Vovchenko:2015vxa}, $K^B_F$ can be defined from $\mu^{*}= \sqrt{{K^B_F}^2+m_B^2 }=\mu^{\textrm{id}}(n_B^{ex},T\rightarrow0)$ without an attraction term. In this work, $K^B_F$ is directly defined from $n_B^{ex}$ as shown in Eq.~\eqref{excdsty}, where the zero-temperature Fermi-Dirac statistics is understood.} In comparison with the parameters introduced in Refs.~\cite{Vovchenko:2015vxa, Zalewski:2015yea, Redlich:2016dpb, Alba:2016hwx, Vovchenko:2017cbu, Vovchenko:2017zpj, Motornenko:2019arp}, higher hard-core density $n_0\simeq 5\rho_0$ will be assumed, as our main object is the hard-core repulsive nature in the high density regime. Thus, the low density properties cannot be accommodated through the parameter set introduced in Refs.~\cite{Vovchenko:2015vxa, Zalewski:2015yea, Redlich:2016dpb, Alba:2016hwx, Vovchenko:2017cbu, Vovchenko:2017zpj, Motornenko:2019arp} even if one considers the attractive contribution appearing in the vdW EoS. For the low density interpolation of this model, one may adopt the various phenomenological potentials developed in previous  literature~\cite{Hebeler:2013nza, Gandolfi:2013baa, Motornenko:2019arp, Kojo:2014rca, Tews:2018kmu, Kojo:2019raj} and refine the model to reproduce the low density properties of nuclear matter. 
For example, the effective model based on the chiral perturbation theory~\cite{Hebeler:2013nza} can be arranged as follows:
\begin{align}
  \epsilon_{B}  \simeq & ~  \frac{ (3 \pi^2)^{\frac{5}{3}}}{10\pi^2 m_{N} } \left(\frac{n_{n}^{\frac{5}{3}}}{ \left(1- n_{\tilde{B}} /n_0 \right)^{\frac{2}{3}} } +\frac{n_{p}^{\frac{5}{3}}}{ \left(1- n_{\tilde{B}} /n_0 \right)^{\frac{2}{3}} } \right)+ m_{N}n_{B} \nonumber\\
  &+ \frac{{k_F^0}^2}{2 m_N}  \rho_0 \Bigg\{ -  \left( (2\sigma -4\sigma_L) \left(\frac{n_p}{n_{B}}\right) \left(\frac{n_n}{n_{B}}
\right) +\sigma_L \right) \left( \frac{n_{B}}{\rho_0}\right)^2  + \left( (2\eta -4 \eta_L) \left(\frac{n_p}{n_B}\right) \left(\frac{n_n}{n_B}\right) +\eta_L \right) \left( \frac{n_{B}}{\rho_0}\right)^{\gamma+1} \Bigg\} \label{ex1},
\end{align}
where $k_F^0=263~\textrm{MeV}$  is the Fermi momentum of nucleons  when $n_B=\rho_0$ and $n_n=n_p$. The first line corresponds to the non-relativistic expansion of the kinetic contributions of nucleons which have intrinsic repulsive nature due to the hard-core size. The trial parameter set $\left\{ \sigma \simeq 7.52,~ \eta \simeq  3.80,~\sigma_L \simeq  1.52,~\eta_L \simeq  0.42,~\gamma \simeq 1.34 \right\}$ reproduces plausible physical properties of the normal nuclear matter: $E/A \simeq -16~\textrm{MeV}$, $e_{sym} \simeq 30~\textrm{MeV}$,~$K_0 \simeq 260~\textrm{MeV}$, and $L_{sym} \simeq 40~\textrm{MeV} $. Nevertheless, only the excluded-volume effect on the baryon system will be considered in this paper to make a simple argument about the role of hard-core repulsion at high density.

An interesting point about this multi flavor hard-core model is the third contribution in the chemical potential~\eqref{hchemp}. Even if there are only few of a specific particle, the chemical potential can be enhanced if the space is taken by the other finite-size particles. This nature can be understood as follows: the excluded volume by some particles makes the system cost more energy to create a fast  particle wave function. The enhanced energy by the strong repulsive mean field at high density is realized in terms of hard-core repulsion with the effective size of the particle.  For the most simplified example, one can guess the flavor symmetric configuration ($\alpha=0$, $m_{i}=m_B$) where $\mu_{n} \simeq \mu_{p}\simeq \mu_{\Lambda}$ are appearing in the hard-core density limit ($n_{\tilde{B}}\simeq n_0$).  However, if an asymmetric configuration is assumed, it is hard to obtain heavier particles having hard-core size even in the high density regime. As an example, the hard-core size of $\Delta$ (1232) can be supposed from the expected repulsive nature in short distance~\cite{Hatsuda:2018nes, Park:2018ukx, Park:2019bsz, Marques:2018mic}. Because its rest mass is heavier than the mass of nucleons by $\approx 300~\textrm{MeV}$ and  most of the system volume is excluded by the other baryons in the high density regime, it is dynamically unfavorable to accommodate the $\Delta$ isobar [see Eq.~\eqref{hchemp}]. If the repulsive core size is evidently smaller than the effective size of other particles, the heavier baryons may emerge around $n_B\sim n_0$.

\begin{figure}
\includegraphics[width=0.95\textwidth]{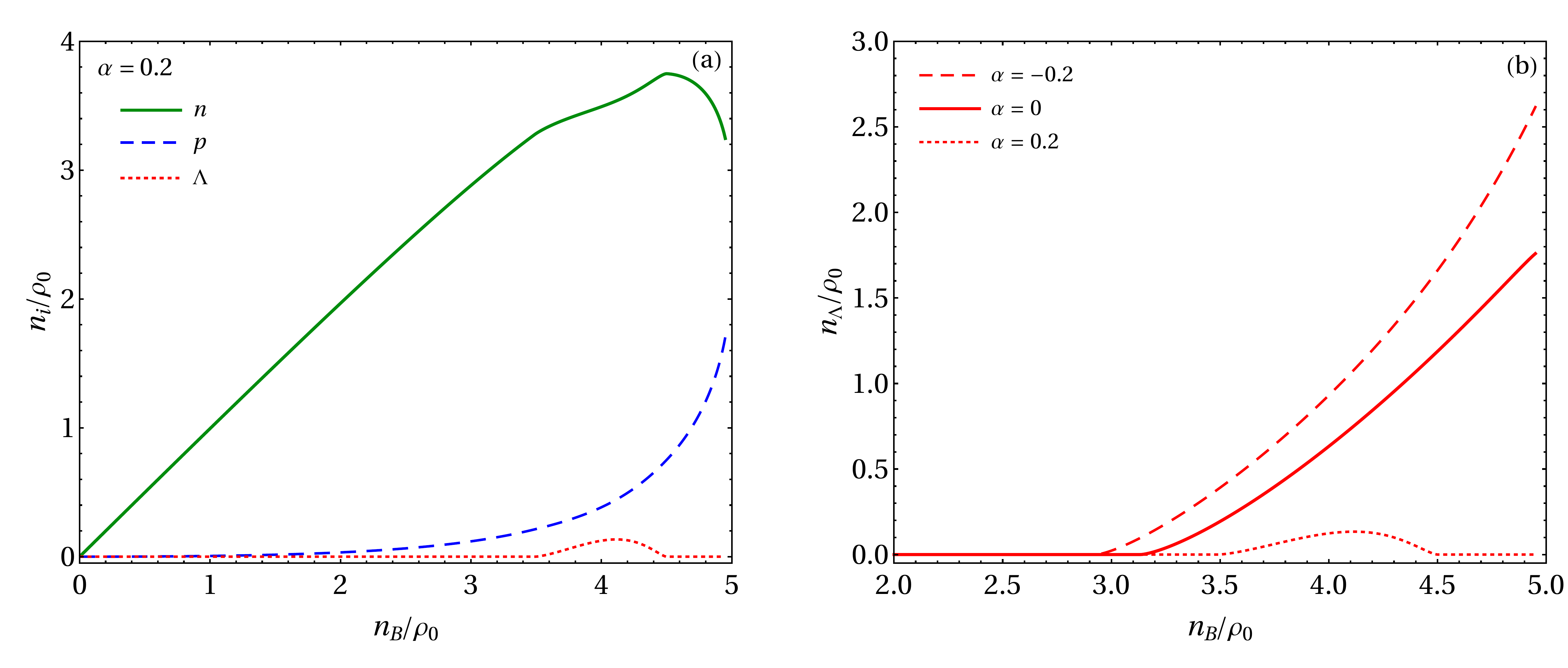}
\caption{Baryon number density within the hard-core repulsion. Green, blue, and red colors represents the densities of neutron, proton, and $\Lambda$ hyperon respectively (left) and the hyperon density with different $\alpha$ parameters for the repulsive core size of the $\Lambda$ hyperon (right).} \label{fig2}
\end{figure}

For the asymmetric realization of three-flavor system, one should consider electromagnetic charge and possible decay channels due to weak interaction: $ n \leftrightarrow p + e$, $ n \leftrightarrow \Lambda$  with $m_{n}=m_{p}\simeq 1~\textrm{GeV}$, $m_{\Lambda} \simeq 1.2~\textrm{GeV}$.  Considering physical constraints,
\begin{align}
n_p&=n_e,\label{cn1}\\
\mu_n&=\mu_p + \mu_e,\label{beq1}\\
\mu_n&=\mu_{\Lambda}~(\textrm{when}~n_{\Lambda}\neq0.~ n_{\Lambda}=0~\textrm{if}~\mu_{\Lambda}<m_{\Lambda}), \label{beq2}
\end{align}
each baryon density can be calculated in $n_{B}$ variation as plotted in Fig.~\ref{fig2}.   As one can find in the density profile plotted in Fig.~\ref{fig2}(a), the large scale of $\alpha$ (stronger repulsion) compared to the other baryons expels the $\Lambda$ hyperon. In this case $ n_{\Lambda}$ vanishes around $n_B \sim n_0$. One can guess the reason for this from the non-relativistic expansion of Eq.~\eqref{hchemp}, where the chemical potential has singular contributions near the hard-core limit. For the $\mu_{\Lambda}$ with $\alpha=0.2$, the contribution of the third term dominates the other contribution but this contribution is weaker in the $\alpha=0$ case. The less singular behavior allows $n_n \simeq n_{\Lambda}$ by~\eqref{beq2}. 


 \begin{figure}
\includegraphics[width=0.48\textwidth]{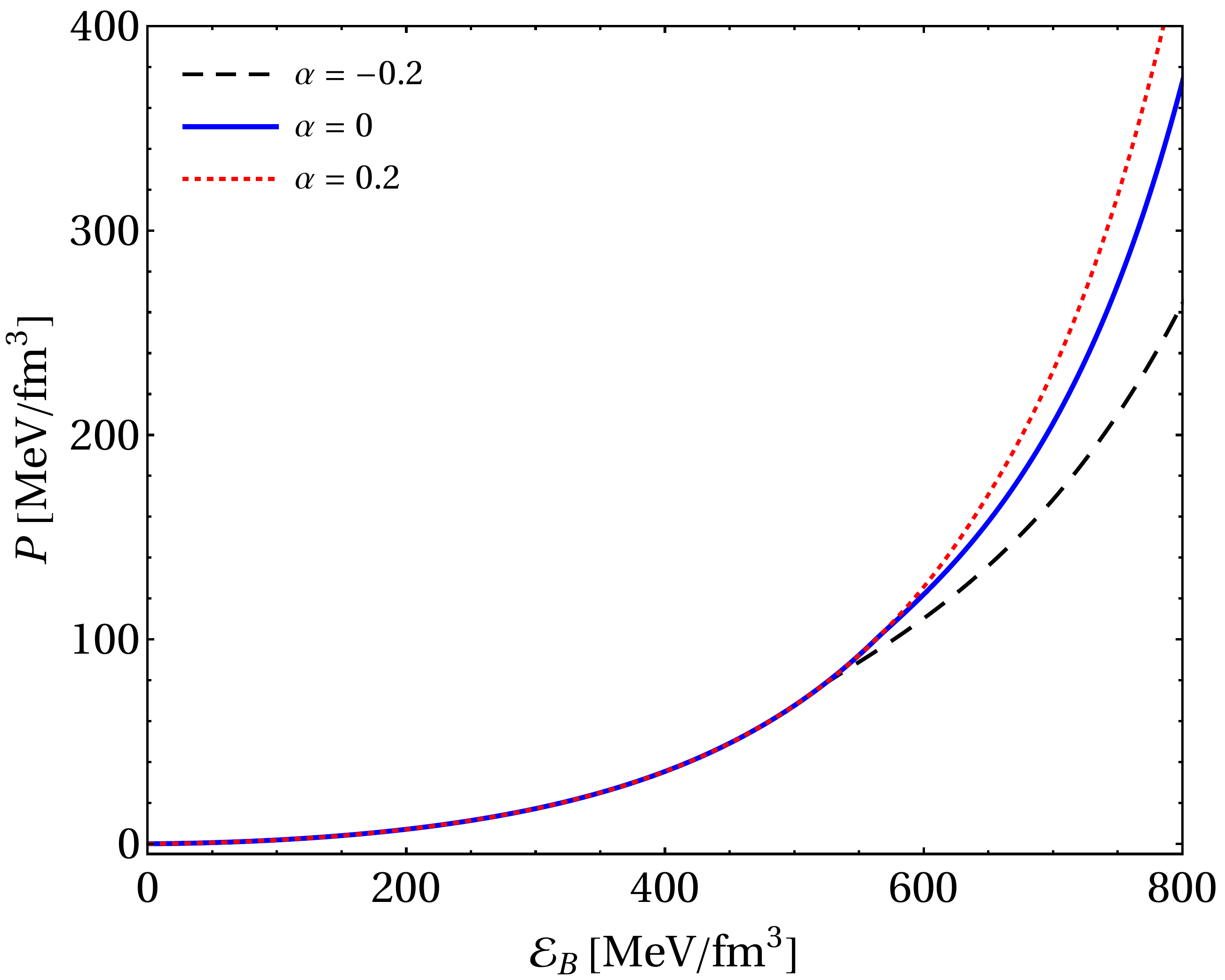}
\caption{Equation of state of baryonic system in hard-core repulsive interaction ($n_0=5\rho_0$).} 
\label{fig3}
\end{figure}

In Fig.~\ref{fig2}(b), the hyperon density $n_{\Lambda}$ is presented for three different values of $\alpha$. In the symmetric hard-core input, $n_{\Lambda} \simeq n_n$ around the hard-core limit, although the rest mass of $\Lambda$ hyperon is higher than nucleon ($m_n=m_p < m_{\Lambda}$). A similar density configuration appears when a weaker repulsion (negative $\alpha$) is assigned. 
The corresponding EoS is plotted in Fig.~\ref{fig3}, for three different values of $\alpha$. We observe that decreasing $\alpha$ or even using a negative value leads to the very stiff EoS, due to the singular hard-core interaction. As one can find in Eq.~\eqref{hchemp}, all chemical potentials have a singularity around the hard-core limit, which leads to causality violating $v_s^2 \gg 1$. At this stage, one can recall the argument of the Hagedorn model~\cite{Hagedorn:1965st} and construct an analogous argument. According to the Hagedorn model, extensive values such as entropy have a singularity around the critical temperature $S / S_{0} \sim  (T_{0}-T )^{-a}$  if the energy spectrum grows as $\rho(E) \sim e^{E/T_{0}}$. When the temperature exceeds $T_{0}$, the Hagedorn system prefers to create new degrees of freedom rather than keep the exponentially populating high energy states. In our system, the EoS has intrinsic singularity around the hard-core density. If the total baryon number density goes beyond the hard-core density, the system would prefer to generate new degrees of freedom, the quarks.\footnote{By definition~\eqref{excdsty}, the baryon density has upper limit $n_0$ in our excluded-volume model. The argument is given in the context of the repulsive singular potential, without assuming the hard-core size.}

\subsection{Dynamically generated quark degrees of freedom}

    Quark degrees of freedom can be generated via various physical scenarios. If some type of phase transition occurs, the entire nuclear matter becomes quark phase after some critical density. However, the quarkyonic-like model discussed in this paper does not have this kind of nature. In the $N_c \rightarrow \infty$ limit, quark wave functions near the Fermi surface are confined in baryon-like states. In the low density regime, the matter behaves as normal nuclear matter but the quark wave functions become saturated from the low momentum state when the matter density reaches a few times  $\rho_0$, where $k_F^{B} \sim O(\Lambda_{\textrm{QCD}})$. When the saturated quarks form their own Fermi sphere, the momenta of confined wave functions are enhanced by Pauli's exclusion principle as they should take larger momentum than the fully occupied lower phases. Around the onset moment, the pressure is continuous and stiffly increasing as the chemical potential should show stiffness ($k_{F}^{B}\simeq N_c k_{F}^{Q}$), different from the first-order phase transition. Thus, if one models this quarkyonic picture within the hard-core repulsion between the baryons, the appearance of quark degrees of freedom should not accompany any sign of discontinuity of EoS. As a first step, we allow the quark degree of freedom pretending that Pauli's exclusion principle does not exist. The energy density can be written as
\begin{align}
\varepsilon_{\textrm{mix.}} &= \left( 1-  \frac{n_{\tilde{B}}}{n_0} \right) \frac{1}{ \pi^2} \sum_{i }^{\{n,p,\Lambda\}} \int^{K_{F}^{B_i}}_{0} dk k^2  \left(k^2+m_{B_i}^2 \right)^{\frac{1}{2}}+ \frac{ N_c}{\pi^2}   \sum_{j}^{\{u,d,s \}}\int^{k^{Q_j}_F}_0  d k k^2  \left(k^2+m_{Q_j}^2 \right)^{\frac{1}{2}}+ \frac{(3 \pi^2)^{\frac{4}{3}}}{4\pi^2}n_{e}^{\frac{4}{3}}. \label{emixture}
\end{align}
Note that the quark phase is not modified as was done in Ref.~\cite{Jeong:2019lhv} because Pauli's exclusion principle is not considered yet.\footnote {An artificial singularity can appear when one assigns the explicit shell-like distribution according to Pauli's exclusion principle. The singularity of the baryon side around $n_B\simeq n_0$ can cause noise which violates the total baryon number conservation ($n_{B}= n_{\tilde{B}}+n_{\tilde{Q}} $) in the numerical calculation. By introducing a kind of regulator in the quark phase measure~\cite{Jeong:2019lhv}, the physical solution can be ensured.} Recall the isospin symmetric configuration in Ref.~\cite{Jeong:2019lhv}. If electromagnetic charge is not concerned, the quark degrees of freedom naturally appear by the dynamical requirement $\mu_{B}= N_c \mu_{Q}$. In the three-flavor extension, this simple relation needs modification. Our system described by Eq.~\eqref{emixture} is mixture of baryons and possible quarks with electron clouds. For a given total baryon number density, we have two constraints for the baryon and lepton numbers. The first constraint is obtained from the baryon number conservation:
\begin{align}
n_{B}&= n_{\tilde{B}}+n_{\tilde{Q}},\\
dn_{B}&= dn_{\tilde{B}}+dn_{\tilde{Q}}\nonumber\\
& = dn_n+dn_p+dn_{\Lambda}+dn_{\tilde{u}}+dn_{\tilde{d}}+dn_{\tilde{s}} =0,\label{bc}
\end{align}
where the tilde denotes the unit of baryon number. The second constraint is the  electromagnetic charge neutrality~\eqref{cn1}:
\begin{align}
n_{e}&= n_p+ 2n_{\tilde{u}}-n_{\tilde{d}}-n_{\tilde{s}},\\
dn_{e}&= dn_p+2dn_{\tilde{u}}-dn_{\tilde{d}}-dn_{\tilde{s}}.\label{cn2}
\end{align}
Also, we have the additional equilibrium constraints from the the possible weak-decays:
\begin{align}
\mu_{\tilde{d}}&=\mu_{\tilde{u}}+3\mu_e,\label{beq3}\\
\mu_{\tilde{d}}&=\mu_{\tilde{s}}~(\textrm{when}~n_{\tilde{s}}\neq0.~n_{\tilde{s}}=0~\textrm{if}~\mu_{\tilde{s}}<N_c m_{s}),
\end{align}
where $\mu_{\tilde{Q}_i}=N_c \mu_{Q_i}$ denotes the quark chemical potential in the unit of baryon number.
At the minimum, the deviation of energy density is zero:
\begin{align}
d \varepsilon_{\textrm{mix.}} & = \mu_n dn_n + \mu_p dn_p + \mu_{\Lambda} dn_{\Lambda} + \mu_{\tilde{u}} dn_{\tilde{u}}+ \mu_{\tilde{d}} dn_{\tilde{d}}+ \mu_{\tilde{s}} dn_{\tilde{s}}+ \mu_{e} dn_{e}\nonumber\\
& = \mu_n (dn_n+dn_p) + \mu_{\Lambda} dn_{\Lambda} +  \left( \mu_{\tilde{d}} -\mu_e \right) (dn_{\tilde{u}}+dn_{\tilde{d}}) + \left( \mu_{\tilde{s}} -\mu_e \right) dn_{\tilde{s}}=0,\label{emin}
\end{align}
where the aforementioned constraints are used. By using the constraint~\eqref{bc}, one can obtain following relation:
\begin{align}
\mu_n&= N_c \mu_{d} -\mu_e,\label{minc0}
\end{align}
 which corresponds to the three-flavor modification of the constraint $\mu_N = N_c \mu_q$ in Ref.~\cite{Jeong:2019lhv}. This constraint has the following conditions in each different configuration:
\begin{align}
\textrm{if}~n_{\Lambda}&=0,~n_{s}=0,~ \mu_n= N_c \mu_{d} -\mu_e =\mu_p + \mu_e, \label{minc1} \\
\textrm{if}~n_{\Lambda}&\neq0,~n_{s}\neq0,~ \mu_n= N_c \mu_{d} -\mu_e = \mu_{\Lambda}=\mu_p + \mu_e= N_c \mu_{s} -\mu_e,\label{minc2}\\
\textrm{if}~n_{\Lambda} &= 0,~n_{s}\neq0,~ \mu_n= N_c \mu_{d} -\mu_e = \mu_{\Lambda}=\mu_p + \mu_e,\label{minc3}\\
\textrm{if}~n_{\Lambda}&\neq0,~n_{s}=0,~ \mu_n= N_c \mu_{d} -\mu_e =\mu_p + \mu_e= N_c \mu_{s} -\mu_e.\label{minc4}
\end{align}

\begin{figure}
\includegraphics[width=0.95\textwidth]{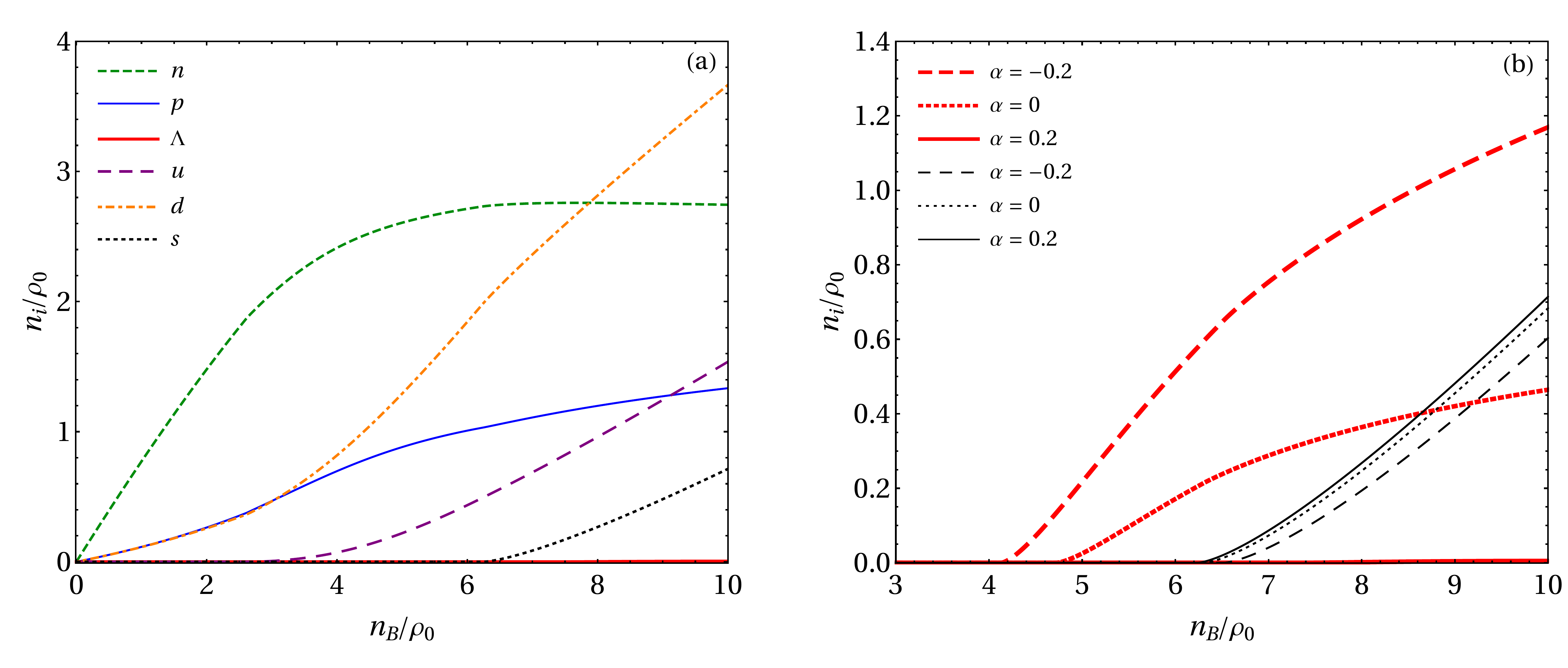}
\caption{Density profile for $\alpha=0.2$ (a) and strangeness density profile for different values of $\alpha$ (b). Density profile (a): green short-dashed line, blue solid line (thin), and red solid line (thick) (corresponding to $n_{\Lambda}=0$) represent the density profiles of neutron, proton, and $\Lambda$ hyperon, respectively. Purple dashed line, orange dot-dashed line, and black dotted line represent the density profiles of $u$, $d$, and $s$ quarks, respectively. Strangeness density profile (b): thick (thin) curves represent $n_{\Lambda}  (n_s)$.} 
\label{fig4}
\end{figure}

Now one can calculate the number density of each particle which satisfies Eq.~\eqref{minc0}, with the baryon number chemical potential and the stiffness of the EoS  determined from the density configuration. In Fig.~\ref{fig4}(a) we show the density profile for the mean-field mixture, for $\alpha = 0.2$. The sum of baryon number density is saturated near $n_0$ and $\Lambda$ appears around $n_B \simeq 6\rho_0$ with very small scale. For the hadron sector, it is almost a two-flavor system in the physically relevant regime. On the other hand, in both the symmetric configuration ($\alpha = 0$) and negative $\alpha$, $\Lambda$ appear in slightly lower density regimes $n_B \simeq 4.8\rho_0$ and $n_B \simeq 4.2\rho_0$ with non-negligible amount, as can be seen in Fig.~\ref{fig4}(b). The quark density profile is barely changed but the hadron sector is changed: strong (and positive) $\alpha$ increases the isospin density of the system.  
The $d$ quark degree of freedom appears from  $n_B \simeq 0$ as $\beta$-equilibrium conditions~\eqref{beq1} and \eqref{beq3} are satisfied in the low density regime. Comparing with the plots of Fig.~\ref{fig2}, $d$ quark plays some part of the proton role even in the low density regime. The $\Lambda$ degree of freedom appears in the high density regime, contrary to the early appearance of  the $d$ quark [Eq.~\eqref{beq2} is satisfied at beyond hard-core density as strong $\alpha=0.2$ is assigned]. Because this model has only hard-core repulsive interaction at high density, and does not have any attractive interaction in the low density regime, the density behavior is similar to that of the three-flavor free baryon gas, where $\mu_{\Lambda}>\mu_{n,p}$ in the low density regime. The particle emergence order is not changed by variation of $n_0$ around $5\rho_0$.

\begin{figure}
\includegraphics[width=0.95\textwidth]{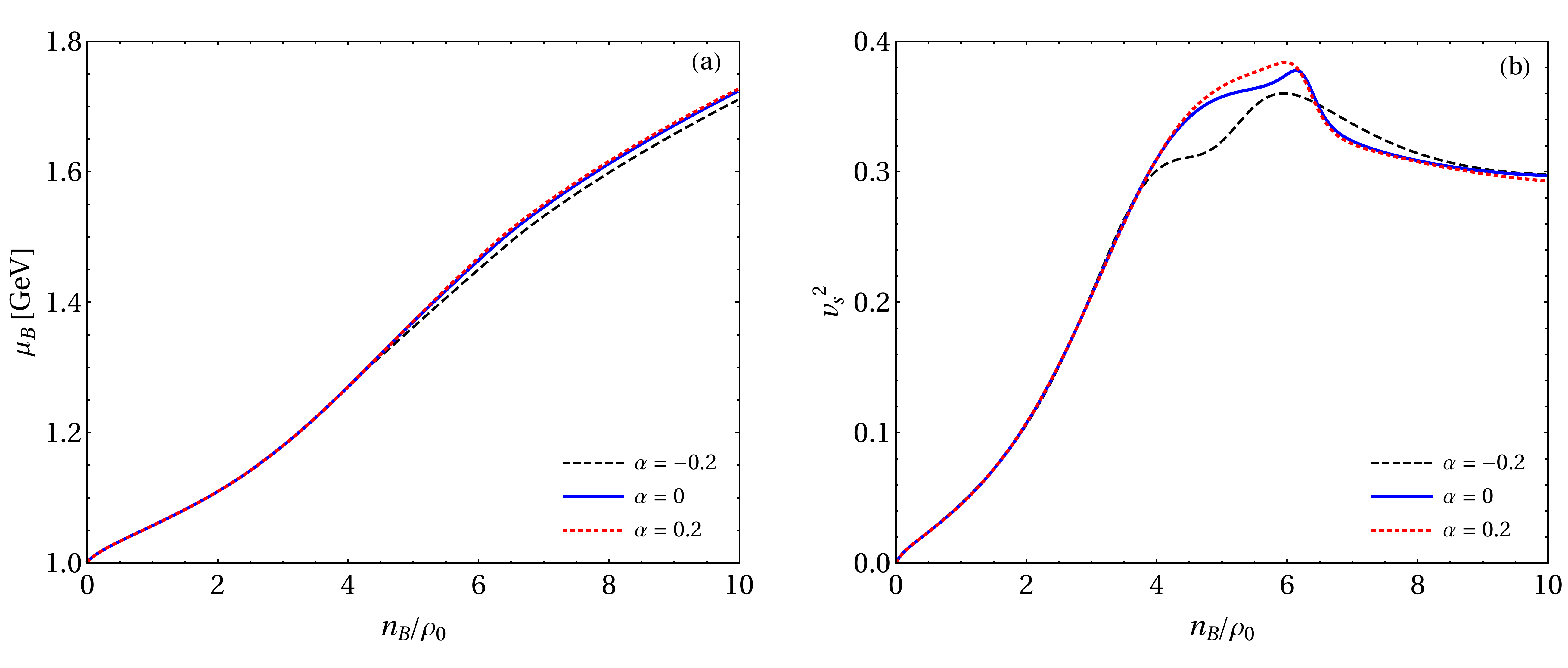}
\caption{Baryon number chemical potential (a) and sound velocity squared (b) with varying the total baryon number.} 
\label{fig5}
\end{figure}

One can check whether the EoS is hard or soft via sound velocity:
\begin{align}
v_s^2=\frac{\partial p}{\partial \varepsilon_{\textrm{mix.}} }= \frac{n_{B}}{\mu_{B} \frac{\partial n_B}{\partial \mu_B}}.
\end{align} 
As one can find in Fig.~\ref{fig5}(a), the baryon number chemical potential shows stiff increment whence $u$ quark degrees of freedom appear. At the moment, $n_{\tilde{B}}$ is already reached around the hard-core density $n_0$. The stiffness becomes moderated at the onset of $s$ quark degrees of freedom. The corresponding $v_s^2$ plotted in Fig.~\ref{fig5}(b) increases to $v_s^2 \simeq 0.4$ around the hard-core regime and seems converge to the ideal limit $1/3$ at the  high density limit. This analysis is valid for three values of $\alpha$ used in this work. The scale of the sound velocity seems not huge enough to accommodate the recent analyses~\cite{Abbott:2018exr, Fattoyev:2017jql, Annala:2017llu, Vuorinen:2018qzx, Raithel:2018ncd, Most:2018hfd, Tews:2019cap, Tews:2019ioa, Capano:2019eae} based on GW observation~\cite{TheLIGOScientific:2017qsa}.

\begin{figure}
\includegraphics[width=0.95\textwidth]{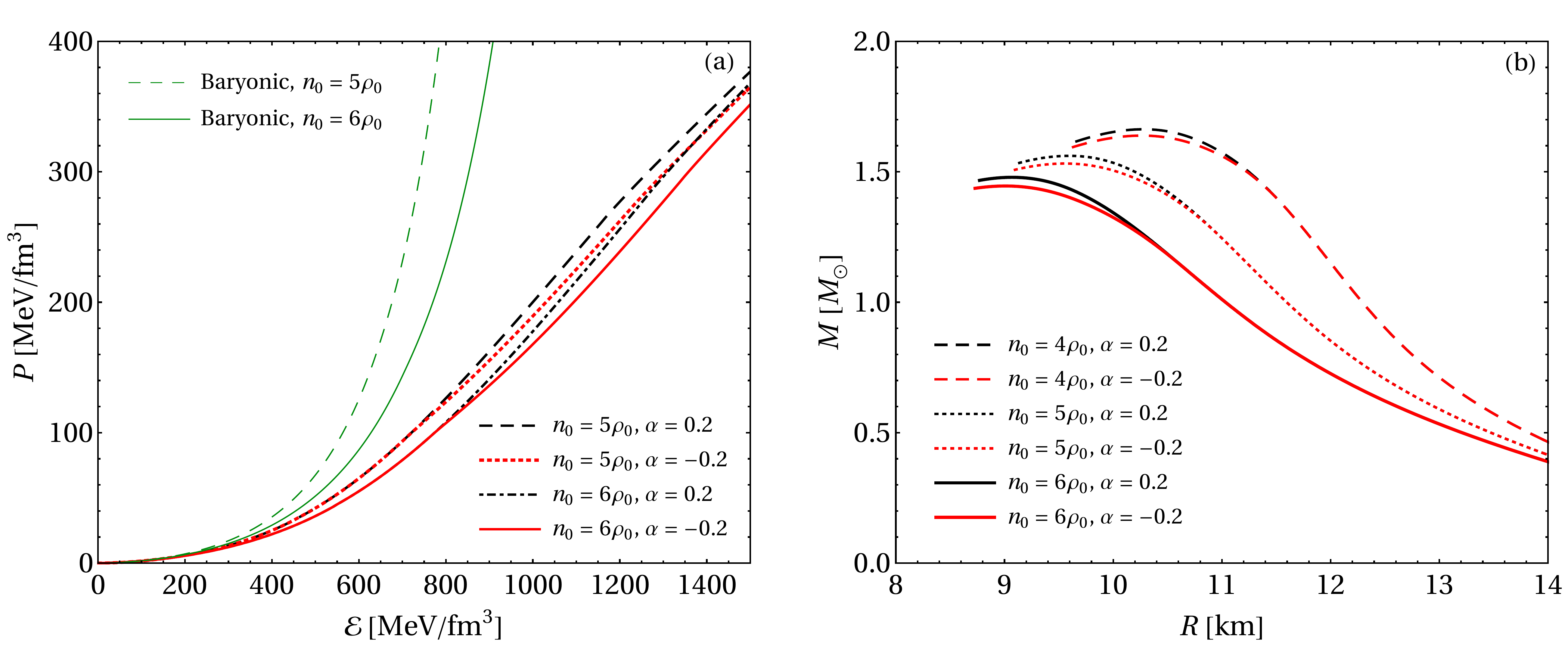}
\caption{Pressure of the baryon-quark mixture system within hard-core repulsive interaction between baryons (a). Green thin dashed and solid curves represent the pressure vs energy density with $n_0 = 5\rho_0 $ and $n_0 = 6\rho_0 $, respectively ($\alpha=0.2$). Mass-radius relations  obtained by solving Tolman–Oppenheimer–Volkoff equations~\cite{Tolman:1939jz, Oppenheimer:1939ne} (b). } 
\label{fig6}
\end{figure}

The variation of the EoS by $\alpha$ and $n_0$ is plotted in Fig.~\ref{fig6}(a).  If one assumes only the baryonic phase, the EoS diverges around $n_B \sim n_0$ and the  physically reasonable energy density (pressure) has its maximal boundary around $n_B < n_0$. On the other hand, the EoS of baryon-quark mixture does not have such a boundary. As one can expect from the sound velocity, the emergence of $\Lambda$ ($\alpha <0$) can make a slightly softer EoS. From the evolution of EoS, one can estimate the maximal neutron star mass by solving Tolman–Oppenheimer–Volkoff (TOV) equations~\cite{Tolman:1939jz, Oppenheimer:1939ne}. The stiffness of the EoS at $n_B > 4\rho_0$ determines the maximum mass of the star, and this behavior can be observed in Fig.~\ref{fig6}(b). As expected from Fig.~\ref{fig6}(a), positive values of $\alpha$ produces slightly larger maximum  mass numbers, while larger values of the hard-core size produces smaller masses.   
However, the current mixture model cannot support the $2 M_{\odot}$ state and the $R_{1.4}\leq 13.5~\textrm{km}$ constraint. The lack of attractive potential leads to too high pressure in the low density regime, and the emergence of quarks at an early stage cannot make a hard enough EoS in the high density regime. Thus, the radius becomes too large at the lower mass state and the higher mass state cannot be obtained as the pressure in the high density regime is not strong enough. Even if the $2 M_{\odot}$ state is obtained by considering an artificially strong repulsive core, $n_0 < 4\rho_0$, the $R_{1.4}\leq 13.5~\textrm{km}$ constraint cannot be satisfied by the same reason. If one considers Pauli's exclusion principle which leads to the shell-like baryon phase distribution, so that a hard enough EoS is obtained around the intermediate density regime, both of the constraints can be satisfied in accordance with the quarkyonic matter concept.

\section{Discussion}\label{sec3}

In this paper, we made a three-flavor extension of the single-flavor excluded-volume model~\cite{Jeong:2019lhv} for the baryon-quark mixture as the first step. This approach  could be understood as the zero-temperature and high density limit of the multi flavor vdW model in Fermi-Dirac statistics~\cite{Yen:1997rv,  Alba:2016hwx,  Vovchenko:2017zpj, Motornenko:2019arp}. Considering electromagnetic charge and possible weak decay channels, the density behavior of each degree of freedom was calculated. The hard-core size of the $\Lambda$ hyperon was parametrized by $\alpha$ whose value would be around $ \vert \alpha \vert \leq 0.2$~\cite{ Nemura:2017bbw, Hatsuda:2018nes, Sasaki:2019qnh}, that leads to an almost two-flavor system which reproduces the hard-soft behavior EoS required by the recent gravitational wave observation.

The stiffness of the EoS is slightly changed by the variation of $\alpha$ for the physically relevant density regime, since the strange particles have relatively small portions [see this behavior in Figs.~\ref{fig4}(b) and \ref{fig5}(b)].
Comparing to the baryonic system, the EoS for the baryon-quark mixture is barely affected by the variation of $\alpha$ [see Figs.~\ref{fig3} and~\ref{fig6}(a)], but among themselves, one can find that the $\alpha = 0.2$ case has a stiffer pressure compared to the other cases because $\Lambda$ degrees of freedom do not appear. This soft EoS leads to the mass-radius relation curve plotted in Fig.~\ref{fig6}(b). The maximal mass is obtained around $1.5M_{\odot}$ and the corresponding radius is smaller than $10~\textrm{km}$ if $n_0 > 5\rho_0$.  One can guess the reason as follows: if Pauli's exclusion principle is not accounted for, the quark degrees of freedom are appearing from the very low density regime ($n_B \simeq 0$). The early onset of the quarks and the lack of attractive interaction lead to a relatively hard EoS in the low density regime ($n_B < \rho_0$). As the constituent quark mass is assumed in this model, the rest mass contribution to the energy density is of similar order to the nucleon contribution while the kinetic contribution is relatively small. Around the hard-core density, the system effectively contains four degrees of freedom, which leads to the relatively soft EoS compared to previously reported studies~\cite{Fattoyev:2017jql, Annala:2017llu, Vuorinen:2018qzx, Raithel:2018ncd, Most:2018hfd, Tews:2019cap, Tews:2019ioa, Capano:2019eae, Masuda:2012ed, Hebeler:2013nza, Gandolfi:2013baa,  Motornenko:2019arp, Kojo:2014rca, Bedaque:2014sqa, Ma:2018qkg, Tews:2018kmu, Kojo:2019raj, Fujimoto:2019hxv}. Therefore, although the hard-core repulsion can enhance the energy density and generates the quark degrees of freedom, so that the plausible tendency is obtained, the current mixture model should be improved in both of the low and high density regime. 

For the low density regime, one may modify the EoS to reproduce the low density properties of nuclear matter. Phenomenological models~\cite{Hebeler:2013nza, Gandolfi:2013baa, Kojo:2014rca, Tews:2018kmu, Kojo:2019raj} can be introduced and refined to satisfy the constraints from the experimental observations as studied in Ref.~\cite{ Motornenko:2019arp}. From the perspective of the excluded-volume approach, one may follow the modifications of the vdW EoS summarized in Ref.~\cite{Vovchenko:2017cbu}, such as Carnahan-Starling modification~\cite{Carnahan}. As a practical approach, a Maxwell construction can be considered between the well constructed EoS in the low density regime and the current approach for the high density regime. In order to make more quantitative predictions to describe the physical EoS in accordance with quarkyonic matter concept, one should consider Pauli's exclusion principle which dynamically generates the shell-like phase distribution for baryons. The nucleons have enhanced momenta by Pauli's exclusion principle ($k_{F}^{B}\simeq N_c k_{F}^{Q}$) in the shell-like momentum distribution, which can lead to a hard enough EoS to support the $2M_{\odot}$ state. In the multi flavor extension of the single-flavor model~\cite{Jeong:2019lhv}, it is expected one will obtain the hard-soft transition  with $v_s^2 > 0.5$ (for $n_B\sim n_0$) and a proper mass-radius curve satisfying $M_{\textrm{max.}} > 2 M_{\odot}$ and $R_{1.4}<13.5~\textrm{km}$, which will be consistent results with the recently reported study~\cite{Motornenko:2019arp}. In this extension, one should consider the baryon number conservation constraint at the onset of the quark Fermi sea and subsequent matching constraints between the perturbative and confined quark momenta, which accompany the complexity in the numerical calculation. 
 Nevertheless, the results of this work will serve as guide for such an  extension, that is now under construction and will be reported elsewhere.

\begin{acknowledgements}
The authors acknowledge useful discussions with Larry McLerran and Sanjay Reddy during development of this work and thank J\'er\^ome Margueron for discussions during his INT visit. The authors also thank Tetsuo Hatsuda and Takashi Inoue for permission to use data and providing useful comments. The authors  acknowledge the support of the Simons Foundation under the Multifarious Minds Program grant 557037. The work of Dyana Duarte, Saul Hernandez-Ortiz and Kie Sang Jeong was supported by the U.S. DOE under Grant No. DE-FG02-00ER41132
\end{acknowledgements}


\begin{thebibliography}{70}


\bibitem{TheLIGOScientific:2017qsa} 
  B.~P.~Abbott {\it et al.} (LIGO Scientific Collaborationand and Virgo Collaboration),
  Phys.\ Rev.\ Lett.\  {\bf 119}, no. 16, 161101 (2017)
  doi:10.1103/PhysRevLett.119.161101
  [arXiv:1710.05832 [gr-qc]].

\bibitem{Abbott:2018exr} 
  B.~P.~Abbott {\it et al.} (LIGO Scientific Collaborationand and Virgo Collaboration),
  Phys.\ Rev.\ Lett.\  {\bf 121}, no. 16, 161101 (2018)
  doi:10.1103/PhysRevLett.121.161101
  [arXiv:1805.11581 [gr-qc]].


\bibitem{Fattoyev:2017jql} 
  F.~J.~Fattoyev, J.~Piekarewicz and C.~J.~Horowitz,
  Phys.\ Rev.\ Lett.\  {\bf 120}, no. 17, 172702 (2018)
  doi:10.1103/PhysRevLett.120.172702
  [arXiv:1711.06615 [nucl-th]].

\bibitem{Annala:2017llu}  
  E.~Annala, T.~Gorda, A.~Kurkela and A.~Vuorinen,
  Phys.\ Rev.\ Lett.\  {\bf 120}, no. 17, 172703 (2018)
  doi:10.1103/PhysRevLett.120.172703
  [arXiv:1711.02644 [astro-ph.HE]].
  
  
\bibitem{Vuorinen:2018qzx} 
  A.~Vuorinen,
  Nucl.\ Phys.\ A {\bf 982}, 36 (2019)
  doi:10.1016/j.nuclphysa.2018.10.011
  [arXiv:1807.04480 [nucl-th]].

\bibitem{Raithel:2018ncd}
  C.~Raithel, F.~Özel and D.~Psaltis,
  Astrophys.\ J.\  {\bf 857}, no. 2, L23 (2018)
  doi:10.3847/2041-8213/aabcbf
  [arXiv:1803.07687 [astro-ph.HE]].
  
  
\bibitem{Most:2018hfd} 
  E.~R.~Most, L.~R.~Weih, L.~Rezzolla and J.~Schaffner-Bielich,
  Phys.\ Rev.\ Lett.\  {\bf 120}, no. 26, 261103 (2018)
  doi:10.1103/PhysRevLett.120.261103
  [arXiv:1803.00549 [gr-qc]].
  
  
\bibitem{Tews:2019cap}
  I.~Tews, J.~Margueron and S.~Reddy,
  Eur.\ Phys.\ J.\ A {\bf 55}, no. 6, 97 (2019)
  doi:10.1140/epja/i2019-12774-6
  [arXiv:1901.09874 [nucl-th]].
  
\bibitem{Tews:2019ioa}
I.~Tews, J.~Margueron and S.~Reddy,
AIP Conf. Proc. \textbf{2127}, no.1, 020009 (2019)
doi:10.1063/1.5117799
[arXiv:1905.11212 [nucl-th]].
  
\bibitem{Capano:2019eae}
C.~D.~Capano, I.~Tews, S.~M.~Brown, B.~Margalit, S.~De, S.~Kumar, D.~A.~Brown, B.~Krishnan and S.~Reddy,
Nature Astron. \textbf{4}, no.6, 625-632 (2020)
doi:10.1038/s41550-020-1014-6
[arXiv:1908.10352 [astro-ph.HE]].
  
  
\bibitem{Demorest}
 P.~B.~Demorest, T.~Pennucci, S.~M.~Ransom, M.~S.~E.~Roberts, and J.~W.~T.~Hessels,
  Nature\ (London)\ {\bf467}, 1081 (2010).
 
\bibitem{Antoniadis} 
 J. Antoniadis et al.,
  Science\  {\bf340}, 1233232 (2013).

  
  
 
\bibitem{Glendenning:1984jr}
  N.~K.~Glendenning,
  Astrophys.\ J.\  {\bf 293}, 470 (1985).


\bibitem{Knorren:1995ds}
  R.~Knorren, M.~Prakash and P.~J.~Ellis,
  Phys.\ Rev.\ C {\bf 52}, 3470 (1995)
  [nucl-th/9506016].
  
  
\bibitem{Brown:1975di}
  G.~E.~Brown and W.~Weise,
  Phys.\ Rep.\  {\bf 22}, 279 (1975).
  doi:10.1016/0370-1573(75)90026-5


\bibitem{Cai:2015hya}
  B.~J.~Cai, F.~J.~Fattoyev, B.~A.~Li and W.~G.~Newton,
  Phys.\ Rev.\ C {\bf 92}, no. 1, 015802 (2015)
  doi:10.1103/PhysRevC.92.015802
  [arXiv:1501.01680 [nucl-th]].
  
\bibitem{Hebeler:2013nza} 
K.~Hebeler, J.~Lattimer, C.~Pethick and A.~Schwenk,
Astrophys. J. \textbf{773}, 11 (2013)
doi:10.1088/0004-637X/773/1/11
[arXiv:1303.4662 [astro-ph.SR]].
   
\bibitem{Gandolfi:2013baa} 
S.~Gandolfi, J.~Carlson, S.~Reddy, A.~Steiner and R.~Wiringa,
Eur. Phys. J. A \textbf{50}, 10 (2014)
doi:10.1140/epja/i2014-14010-5
[arXiv:1307.5815 [nucl-th]].

  
  
\bibitem{McLerran:2007qj}  
  L.~McLerran and R.~D.~Pisarski,
  Nucl.\ Phys.\ A {\bf 796}, 83 (2007)
  doi:10.1016/j.nuclphysa.2007.08.013
  [arXiv:0706.2191 [hep-ph]].
  
  
	
\bibitem{Fukushima:2015bda} 
  K.~Fukushima and T.~Kojo,
  Astrophys.\ J.\  {\bf 817}, no. 2, 180 (2016)
  doi:10.3847/0004-637X/817/2/180
  [arXiv:1509.00356 [nucl-th]].
  
  
 
\bibitem{McLerran:2018hbz}   
  L.~McLerran and S.~Reddy,
  Phys.\ Rev.\ Lett.\  {\bf 122}, no. 12, 122701 (2019)
  doi:10.1103/PhysRevLett.122.122701
  [arXiv:1811.12503 [nucl-th]].
  
 
\bibitem{Jeong:2019lhv}
K.~S.~Jeong, L.~McLerran and S.~Sen,
Phys. Rev. C \textbf{101}, no.3, 035201 (2020)
doi:10.1103/PhysRevC.101.035201
[arXiv:1908.04799 [nucl-th]].
  
  
\bibitem{tHooft:1973alw}  
  G.~'t Hooft,
  Nucl.\ Phys.\ B {\bf 72}, 461 (1974).
  doi:10.1016/0550-3213(74)90154-0
  
  
\bibitem{tHooft:1974pnl} 
  G.~'t Hooft,
  Nucl.\ Phys.\ B {\bf 75}, 461 (1974).
  doi:10.1016/0550-3213(74)90088-1


\bibitem{Hamada:1962nq} 
  T.~Hamada and I.~D.~Johnston,
  Nucl.\ Phys.\  {\bf 34}, 382 (1962).
  doi:10.1016/0029-5582(62)90228-6


\bibitem{Herndon:1967zza} 
  R.~C.~Herndon and Y.~C.~Tang,
  Phys.\ Rev.\  {\bf 153}, 1091 (1967).
  doi:10.1103/PhysRev.153.1091
  
\bibitem{Kurihara:1984mh}  
  Y.~Kurihara, Y.~Akaishi and H.~Tanaka,
  Prog.\ Theor.\ Phys.\  {\bf 71}, 561 (1984).
  doi:10.1143/PTP.71.561
 
\bibitem{Rischke:1991ke} 
D.~H.~Rischke, M.~I.~Gorenstein, H.~Stoecker and W.~Greiner,
Z. Phys. C \textbf{51}, 485-490 (1991)
doi:10.1007/BF01548574
 
\bibitem{Kievsky:1992um} 
  A.~Kievsky, S.~Rosati and M.~Viviani,
  Nucl.\ Phys.\ A {\bf 551}, 241 (1993).
  doi:10.1016/0375-9474(93)90480-L
  
\bibitem{Stoks:1994wp} 
  V.~G.~J.~Stoks, R.~A.~M.~Klomp, C.~P.~F.~Terheggen and J.~J.~de Swart,
  Phys.\ Rev.\ C {\bf 49}, 2950 (1994)
  doi:10.1103/PhysRevC.49.2950
  [nucl-th/9406039].
 
\bibitem{Wiringa:1994wb} 
  R.~B.~Wiringa, V.~G.~J.~Stoks and R.~Schiavilla,
  Phys.\ Rev.\ C {\bf 51} (1995) 38
  doi:10.1103/PhysRevC.51.38
  [nucl-th/9408016].
  
 
  
\bibitem{Yen:1997rv} 
G.~D.~Yen, M.~I.~Gorenstein, W.~Greiner and S.~N.~Yang,
Phys. Rev. C \textbf{56}, 2210-2218 (1997)
doi:10.1103/PhysRevC.56.2210
[arXiv:nucl-th/9711062 [nucl-th]].
  
  
\bibitem{Machleidt:2000ge} 
  R.~Machleidt,
  Phys.\ Rev.\ C {\bf 63}, 024001 (2001)
  doi:10.1103/PhysRevC.63.024001
  [nucl-th/0006014].
  
  
\bibitem{Vovchenko:2015vxa} 
  V.~Vovchenko, D.~V.~Anchishkin and M.~I.~Gorenstein,
  Phys.\ Rev.\ C {\bf 91}, no. 6, 064314 (2015)
  doi:10.1103/PhysRevC.91.064314
  [arXiv:1504.01363 [nucl-th]].
 
  
\bibitem{Zalewski:2015yea}  
  K.~Redlich and K.~Zalewski,
  Phys.\ Rev.\ C {\bf 93}, no. 1, 014910 (2016)
  doi:10.1103/PhysRevC.93.014910
  [arXiv:1507.05433 [hep-ph]].
 
 
\bibitem{Redlich:2016dpb}  
  K.~Redlich and K.~Zalewski,
  Acta Phys.\ Polon.\ B {\bf 47}, 1943 (2016)
  doi:10.5506/APhysPolB.47.1943
  [arXiv:1605.09686 [cond-mat.quant-gas]].

\bibitem{Alba:2016hwx} 
P.~Alba, V.~Vovchenko, M.~Gorenstein and H.~Stoecker,
Nucl. Phys. A \textbf{974}, 22-34 (2018)
doi:10.1016/j.nuclphysa.2018.03.007
[arXiv:1606.06542 [hep-ph]].
  


\bibitem{Vovchenko:2017cbu}  
V.~Vovchenko,
Phys. Rev. C \textbf{96}, no.1, 015206 (2017)
doi:10.1103/PhysRevC.96.015206
[arXiv:1701.06524 [nucl-th]].

  
\bibitem{Vovchenko:2017zpj} 
V.~Vovchenko, A.~Motornenko, P.~Alba, M.~I.~Gorenstein, L.~M.~Satarov and H.~Stoecker,
Phys. Rev. C \textbf{96}, no.4, 045202 (2017)
doi:10.1103/PhysRevC.96.045202
[arXiv:1707.09215 [nucl-th]].

\bibitem{Motornenko:2019arp}
A.~Motornenko, J.~Steinheimer, V.~Vovchenko, S.~Schramm and H.~Stoecker,
Phys. Rev. C \textbf{101}, no.3, 034904 (2020)
doi:10.1103/PhysRevC.101.034904
[arXiv:1905.00866 [hep-ph]].



  
\bibitem{Ishii:2006ec} 
  N.~Ishii, S.~Aoki and T.~Hatsuda,
  Phys.\ Rev.\ Lett.\  {\bf 99}, 022001 (2007)
  doi:10.1103/PhysRevLett.99.022001
  [nucl-th/0611096].
  
\bibitem{Inoue:2016qxt}
T.~Inoue [LATTICE-HALQCD],
PoS \textbf{INPC2016}, 277 (2016)
doi:10.22323/1.281.0277
[arXiv:1612.08399 [hep-lat]].
  
   
\bibitem{Nemura:2017bbw}
H.~Nemura, S.~Aoki, T.~Doi, S.~Gongyo, T.~Hatsuda, Y.~Ikeda, T.~Inoue, T.~Iritani, N.~Ishii, T.~Miyamoto, K.~Murano and K.~Sasaki,
PoS \textbf{LATTICE2016}, 101 (2017)
doi:10.22323/1.256.0101
[arXiv:1702.00734 [hep-lat]].
 
\bibitem{Hatsuda:2018nes}
  T.~Hatsuda,
  Front.\ Phys.\ (Beijing) {\bf 13}, no. 6, 132105 (2018).
  doi:10.1007/s11467-018-0829-4
  
\bibitem{Park:2018ukx}  
  A.~Park, W.~Park and S.~H.~Lee,
  Phys.\ Rev.\ D {\bf 98}, no. 3, 034001 (2018)
  doi:10.1103/PhysRevD.98.034001
  [arXiv:1801.10350 [hep-ph]].
 
\bibitem{Inoue:2018axd}
T.~Inoue [HAL QCD],
AIP Conf. Proc. \textbf{2130}, no.1, 020002 (2019)
doi:10.1063/1.5118370
[arXiv:1809.08932 [hep-lat]].

\bibitem{Park:2019bsz}
A.~Park, S.~H.~Lee, T.~Inoue and T.~Hatsuda,
Eur. Phys. J. A \textbf{56}, no.3, 93 (2020)
doi:10.1140/epja/s10050-020-00078-z
[arXiv:1907.06351 [hep-ph]].
  
\bibitem{Sasaki:2019qnh} 
  K.~Sasaki {\it et al.},
  arXiv:1912.08630 [hep-lat].
  
  

  
\bibitem{Masuda:2012ed}
K.~Masuda, T.~Hatsuda and T.~Takatsuka,
Prog.\ Theor.\ Exp.\ Phys. \textbf{2013}, no.7, 073D01 (2013)
doi:10.1093/ptep/ptt045
[arXiv:1212.6803 [nucl-th]].
   
\bibitem{Kojo:2014rca} 
  T.~Kojo, P.~D.~Powell, Y.~Song and G.~Baym,
  Phys.\ Rev.\ D {\bf 91}, no. 4, 045003 (2015)
  doi:10.1103/PhysRevD.91.045003
  [arXiv:1412.1108 [hep-ph]].
  
\bibitem{Bedaque:2014sqa} 
  P.~Bedaque and A.~W.~Steiner,
  Phys.\ Rev.\ Lett.\  {\bf 114}, no. 3, 031103 (2015)
  doi:10.1103/PhysRevLett.114.031103
  [arXiv:1408.5116 [nucl-th]].

\bibitem{Ma:2018qkg}
Y.~L.~Ma and M.~Rho,
Phys. Rev. D \textbf{100}, no.11, 114003 (2019)
doi:10.1103/PhysRevD.100.114003
[arXiv:1811.07071 [nucl-th]].


\bibitem{Tews:2018kmu} 
  I.~Tews, J.~Carlson, S.~Gandolfi and S.~Reddy,
  Astrophys.\ J.\  {\bf 860}, no. 2, 149 (2018)
  doi:10.3847/1538-4357/aac267
  [arXiv:1801.01923 [nucl-th]].

 
\bibitem{Kojo:2019raj}
T.~Kojo,
AIP Conf. Proc. \textbf{2127}, no.1, 020023 (2019)
doi:10.1063/1.5117813
[arXiv:1904.05080 [astro-ph.HE]].
  
\bibitem{Fujimoto:2019hxv}
Y.~Fujimoto, K.~Fukushima and K.~Murase,
Phys. Rev. D \textbf{101}, no.5, 054016 (2020)
doi:10.1103/PhysRevD.101.054016
[arXiv:1903.03400 [nucl-th]].

\bibitem{Kaplan:1986yq} 
  D.~B.~Kaplan and A.~E.~Nelson,
  Phys.\ Lett.\ B {\bf 175}, 57 (1986).
  doi:10.1016/0370-2693(86)90331-X



\bibitem{Savage:1995kv} 
  M.~J.~Savage and M.~B.~Wise,
  Phys.\ Rev.\ D {\bf 53}, 349 (1996)
  doi:10.1103/PhysRevD.53.349
  [hep-ph/9507288].
  
\bibitem{Jeong:2016qlk} 
  K.~S.~Jeong, G.~Gye and S.~H.~Lee,
  Phys.\ Rev.\ C {\bf 94}, no. 6, 065201 (2016)
  doi:10.1103/PhysRevC.94.065201
  [arXiv:1606.00594 [nucl-th]].
  
\bibitem{Marques:2018mic}
J.~Marques L., S.~H.~Lee, A.~Park, R.~D.~Matheus and K.~S.~Jeong,
Phys. Rev. C \textbf{98}, no.2, 025206 (2018)
doi:10.1103/PhysRevC.98.025206
[arXiv:1806.01668 [nucl-th]].


\bibitem{Hagedorn:1965st}
R.~Hagedorn,
Nuovo Cimento Suppl. \textbf{3}, 147-186 (1965)
CERN-TH-520.



\bibitem{Tolman:1939jz} 
R.~C.~Tolman,
Phys. Rev. \textbf{55}, 364-373 (1939)
doi:10.1103/PhysRev.55.364

\bibitem{Oppenheimer:1939ne}
J.~Oppenheimer and G.~Volkoff,
Phys. Rev. \textbf{55}, 374-381 (1939)
doi:10.1103/PhysRev.55.374






\bibitem{Carnahan}
N.~F.~Carnahan and K.~E.~Starling,
J. Chem. Phys. \textbf{51}, 635 (1969)
doi:10.1063/1.1672048








\end{thebibliography}
\end{document}